\documentclass[aps,prb,reprint,superscriptaddress]{revtex4-2}
\usepackage{physics}
\usepackage{bm}
\usepackage{graphicx}
\usepackage{xcolor}
\usepackage{amsmath,amssymb}

\usepackage{hyperref}
\hypersetup{
	citecolor = blue,
	colorlinks = true,
	urlcolor = blue
}

\allowdisplaybreaks[4]

\begin{document}

%%%%%%%%%% -------------------------------------------------- %%%%%%%%%%

\title{Microscopic Theory of Nonlinear Hall Effect Induced by Electric Field and Temperature Gradient}
\author{Terufumi Yamaguchi}
\affiliation{RIKEN Center for Emergent Matter Science, 2-1 Hirosawa, Wako, Saitama 351-0198, Japan}
\author{Kazuki Nakazawa}
\affiliation{Department of Applied Physics, University of Tokyo, Tokyo 113-8656, Japan}
\author{Ai Yamakage}
\affiliation{Department of Physics, Nagoya University, Nagoya 464-8602, Japan}
\date{\today}

%%%%%%%%%% -------------------------------------------------- %%%%%%%%%%

%%%%%%%%%% -------------------------------------------------- %%%%%%%%%%

\begin{abstract}
	Electric current flows parallel to the outer product of an applied electric field and temperature gradient, 
	a phenomenon we call the nonlinear chiral thermo-electric (NCTE) Hall effect. 
	We present a general microscopic formulation of this effect and demonstrate its existence in a chiral crystal. 
	We show that the contribution of the orbital magnetic moment, which has been previously overlooked, 
	is just as significant as the conventional Berry curvature dipole term. 
	Furthermore, we demonstrate a substantial NCTE Hall effect in a chiral Weyl semimetal. 
	These findings offer new insights into nonlinear transport phenomena and have significant implications 
	for the field of condensed matter physics.
\end{abstract}
\maketitle

%%%%%%%%%% -------------------------------------------------- %%%%%%%%%%

%%%%%%%%%% -------------------------------------------------- %%%%%%%%%%

%%%%% ----- Figure ----- %%%%%
\begin{figure}[t]
	\begin{center}
        \includegraphics[width=8cm]{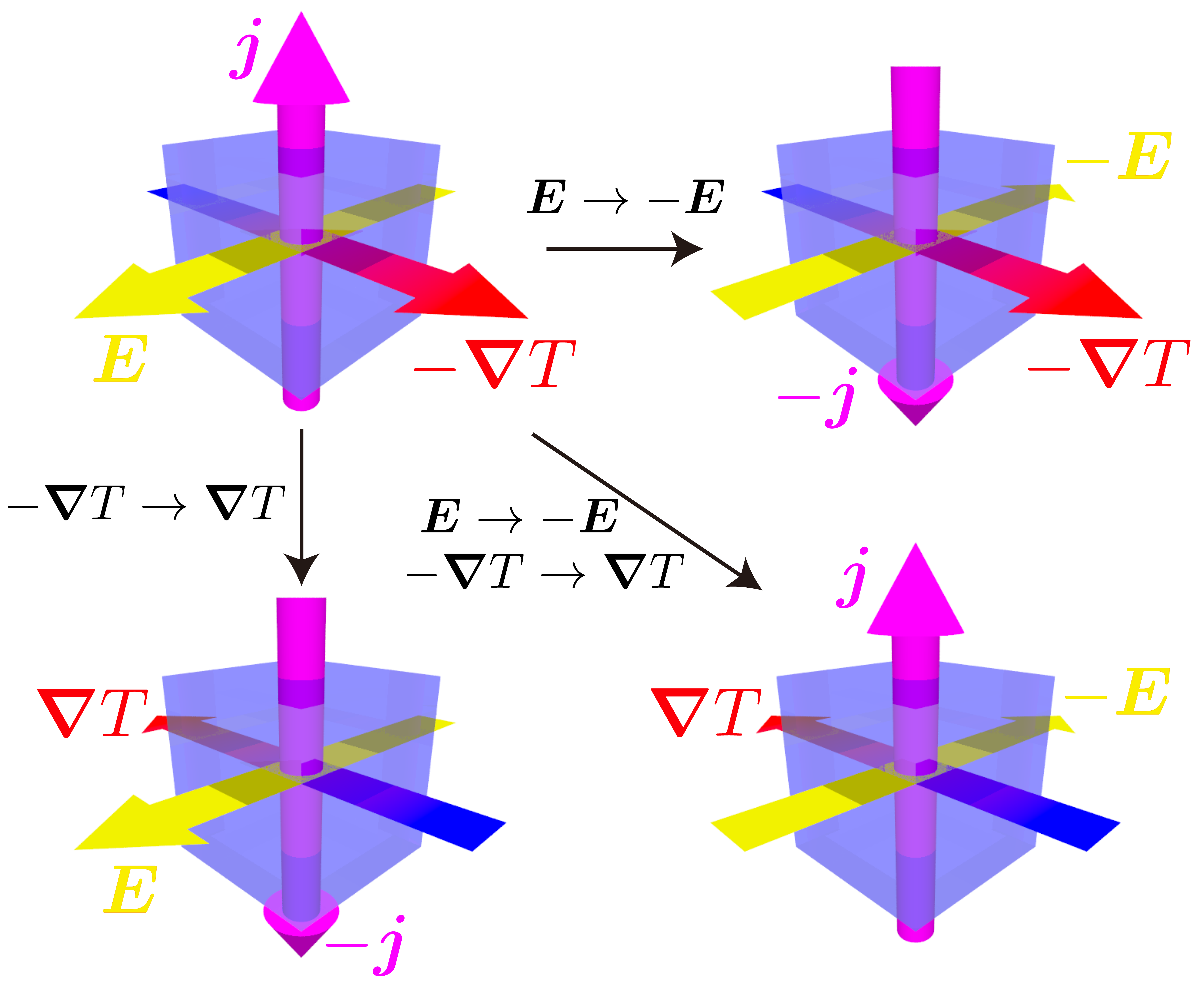}
		\caption{
			Conceptual figure of the NCTE Hall effect. When applying an electric field $\bm{E}$
			and a temperature gradient $- \bm{\nabla} T$ orthogonal to each other, 
			a current $\bm{j}$ flows perpendicular to both of them, $\bm{j} \propto \bm{E} \times (- \bm{\nabla} T)$. 
			Since the NCTE Hall effect is the nonlinear response, the sign of the current changes 
			when replacing $\bm{E} \to - \bm{E}$ or $- \bm{\nabla} T \to \bm{\nabla}T$, 
			whereas the sign is the same when replacing both $\bm{E} \to - \bm{E}$ and $- \bm{\nabla} T \to \bm{\nabla} T$.
			\vspace{-3ex}}
		\label{fig:NCTE}
	\end{center}
\end{figure}
%%%%% ----- END Figure ----- %%%%%

\textit{Introduction.--}
Understanding quantum transport phenomena is essential in physics, 
as it allows us to determine the physical properties by observing how physical quantities respond to different external forces.
We can also design and control materials and their structures based on the information on the transport coefficients 
for high-performance device applications.
A typical example is the response of an electric current to an electric field; 
the Ohm's law and the Hall effect~\cite{Hall1879} are historically well-known, 
and the anomalous Hall effect~\cite{Sinitsyn2008,Nagaosa2010a,Chang2013a} 
and the topological Hall effect~\cite{Bruno2004a,Tatara2002,Neubauer2009,Nakazawa2018} have been extensively studied in recent years. 
Employing a temperature gradient instead of an electric field, these phenomena are known as
the Nernst effect and 
the anomalous (topological) Nernst effect~\cite{Suryanarayanan1999,Lee2004,Hanasaki2008-lw,Mizuguchi2019,Shiomi2013,Mizuta2016}.
These transport phenomena have been mainly studied as linear responses to external forces.
On the other hand, there are responses to two or more external fields, namely nonlinear responses. 
Recently, nonlinear responses have been extensively studied, 
such as nonlinear optical responses, nonreciprocal transport, 
and the nonlinear Hall(Nernst) effect~\cite{Sodemann2015,Takashima2018-ks,Parker2019,Ma2019-wh,Ma2019-wh,Ishizuka2020-uu,Michishita2021,Du2021-tv,Du2021-nc,Zeng2022-tx,Zeng2022-za}.

Here, we focus on the response of an electric current to the outer product of an electric field and a temperature gradient, 
which we call the nonlinear chiral thermo-electric (NCTE) Hall effect. 
The NCTE Hall effect is different from the superposition of linear responses; 
the direction of the NCTE Hall current changes when a direction of either an electric field or temperature gradient changes, 
whereas the direction does not change when the direction of both external forces change (see Fig.~\ref{fig:NCTE}). 
In other words, reversing the sign of one of the two ``inputs" reverses the sign of the ``output",
and reversing the sign of both ``inputs" does not change the sign of the ``output"; that is, it works the same as an XOR logic circuit.

The existence of the NCTE Hall effect has been predicted in Weyl fermion systems~\cite{Hidaka2018}, 
and the description of the Berry curvature dipole has been obtained within semiclassical kinetic theory~\cite{Nakai2019,Toshio2020}.
These studies have shown only in Weyl systems and diverge at the low-temperature limit 
or have only pointed out the possibility of the NCTE Hall effect. 
The microscopic formulation of the NCTE Hall effect for general band structures, 
verifying the finite NCTE Hall conductivities in concrete models and showing the NCTE Hall effect in actual crystals is still absent.

This letter clarifies that the NCTE Hall effect occurs in chiral crystals, based on the microscopic theory we developed. 
First, we formulate the NCTE Hall effect microscopically by employing nonequilibrium (Keldysh) Green's functions~\cite{Kamenev} for the (nonlinear) responses not only to mechanical forces but also to the statistical forces.  
Next, by rewriting our formula in band representation within the relaxation time approximation, 
we find the novel terms expressed in the orbital magnetic moment adding to the conventional Berry curvature dipole terms.
Applying our formula to a minimal model, 
we unveil the finite NCTE Hall conductivity, in which the orbital magnetic moment terms are essential.
Finally, we demonstrate the finite NCTE Hall conductivity in a model of chiral crystal proposed in Ref.~\cite{Yoda2015,Yoda2018}, 
and obtain the NCTE Hall conductivity with experimentally measurable value.

%%%%%%%%%% -------------------------------------------------- %%%%%%%%%%

%%%%%%%%%% -------------------------------------------------- %%%%%%%%%%

\textit{Formulation of the nonlinear chiral thermo-electric Hall effect.--}
We consider the Hamiltonian described as
\begin{align}
    \mathcal{H}
    &=
    \mathcal{H}_0 + \mathcal{H}_{\mathrm{imp}} + \mathcal{H}_{\mathrm{ext}},
    \\
    \mathcal{H}_0
    &=
    \sum_{\bm{k}}
    c^{\dag}_{\bm{k}} H_{\bm{k}} c_{\bm{k}},
    \\
    \mathcal{H}_{\mathrm{imp}}
    &=
    \int \mathrm{d} \bm{r}
    c^{\dag}(\bm{r}) V_{\mathrm{imp}} (\bm{r}) c(\bm{r}),
    \\
    \mathcal{H}_{\mathrm{ext}}
    &=
    e \int \mathrm{d} \bm{r}
    \hat{n}(\bm{r}) \phi(\bm{r},t),
\end{align}
where $c^{\dag}$ $(c)$ is the creation (annihilation) operator,
$V_{\mathrm{imp}}(\bm{r})$ is the impurity potential, 
$e<0$ is an electron charge,
$\hat{n}(\bm{r}) = c^{\dag}(\bm{r}) c(\bm{r})$ is the number density operator, 
and $\phi(\bm{r},t)$ is the scalar potential.
Note that $H_{\bm{k}}$ is, in general, the spin and orbital space matrix.
We define the retarded Green's function as
\begin{align}
	G^{\mathrm{R}}_{\bm{k}}(\varepsilon)
	=
	\qty[
		\varepsilon - H_{\bm{k}} - \Sigma^{\mathrm{R}}_{\bm{k}}(\varepsilon)
	]^{-1},
	\label{eq:GR}
\end{align}
where $\Sigma^{\mathrm{R}}_{\bm{k}}(\varepsilon)$ is the retarded self-energy
due to the impurity scattering. 
We take the impurity average to retain the translational symmetry.
The expectation value of physical quantity $\hat{O}$ can be calculated by using the Keldysh Green's function as
\begin{align}
	\langle \hat{O} \rangle
	=
	- \frac{i}{2} \mathrm{tr} \left[ \hat{O} \star G^{\mathrm{K}} \right].
	\label{eq:ev_GK}
\end{align}
where the product $\star$ is the star product
which is equivalent to convolution integral in the real time-space representation,
or Moyal product in the Wigner representation.
To obtain the Keldysh Green's function, we calculate the Keldysh component of the Dyson's equation,
\begin{align}
	G^{\mathrm{K}}
	=
	G^{\mathrm{R}} \star \Sigma^{\mathrm{K}} \star G^{\mathrm{A}},
	\label{eq:Dyson}
\end{align}
with the self-energy of the Keldysh component whose distribution function is described as local equilibrium,
\begin{align}
	& \Sigma^{\mathrm{K}}
	=
	\qty( 1 - 2 f_{\mathrm{leq}} \qty[\varepsilon; \mu, T(\bm{r})] )
	\qty(\Sigma^{\mathrm{R}} - \Sigma^{\mathrm{A}}),
	\label{eq:Sigma_K}
	\\
	&f_{\mathrm{leq}} \qty[\varepsilon ; \mu, T(\bm{r})]
	=
	\frac{1}{\exp\qty[(\varepsilon - \mu)/T(\bm{r})] + 1 }.
	\label{eq:f_local}
\end{align}
The response to the temperature gradient is obtained by taking the terms $\nabla T(\bm{r})$
from the Dyson's equation (\ref{eq:Dyson}).
In linear response theory, the response to the temperature gradient is often calculated by introducing the gravitational potential
\cite{Luttinger1964}.
The essence of this method is that one calculates the response to a gradient of gravitational potential based on the Kubo formula,
and then replaces the gradient of gravitational potential with the temperature gradient for the nonequilibrium component 
using the Einstein-Luttinger relation.
However, we note that the Einstein-Luttinger relation is only applicable near equilibrium states, 
namely, in the linear response regime.
In other words, such a replacement is not justified in nonlinear response regimes.
In fact, violations of the Einstein-Luttinger relation in specific cases have been reported~\cite{Park2022}.
On the other hand, nonlinear responses to the temperature gradient are calculated based on the Boltzmann equations
using the local equilibrium distribution function as the initial condition~\cite{Takashima2018-ks,Nakazawa2022}.
The method used in this letter is a similar manner but treats fully quantum mechanical way.
We also incorporate an electric field by treating it in Keldysh space to obtain the nonlinear response to a temperature gradient
and an electric field.
For the latter convenience, we consider the setup in which the electric field and temperature gradient are applied to $xy$ plane.
The NCTE Hall current $j_{z}$ is expressed as
\begin{align}
	j_{z} = \sigma^{\mathrm{NCTE}}_{z}
	\qty[
		\bm{E} \times \left( - \frac{\bm{\nabla} T}{T} \right)
	]_{z},
	\label{eq:NCTE_conductivity}
\end{align}
with the NCTE Hall conductivity $\sigma^{\mathrm{NCTE}}_{z}$ 
\begin{align}
	\sigma^{\mathrm{NCTE}}_{z} = &
	-\frac{e^{2}}{4 \pi} 
	\int \mathrm{d} \varepsilon
	\qty( - \frac{\partial f}{\partial \varepsilon} ) (\varepsilon - \mu)
	\sum_{\bm{k}}
	\notag \\
	& \times
	\Im
	\left\{
		\tr
		\left[
			v_{z} G^{\mathrm{R}} v_{x} (\partial_{\varepsilon} G^{\mathrm{R}}) v_{y} G^{\mathrm{R}}
		\right.
	\right.
	\notag \\
	& %\quad
	\left.
		\left.
			+
			v_{z} (\partial_{\varepsilon} G^{\mathrm{R}}) v_{x} G^{\mathrm{R}} v_{y} G^{\mathrm{A}}
			-
			v_{z} G^{\mathrm{R}} v_{x} (\partial_{\varepsilon} G^{\mathrm{R}}) v_{y} G^{\mathrm{A}}
		\right.
	\right.
	\notag \\
	& %\quad
	\left.
		\left.
			-
			v_{z} G^{\mathrm{R}} v_{x} G^{\mathrm{R}} v_{y} (\partial_{\varepsilon} G^{\mathrm{A}})
		\right]
		- 
		\left( v_{x} \leftrightarrow v_{y} \right)
	\right\},
	\label{eq:NCTE_general}
\end{align}
where $v_{i} = \partial_{k_{i}} H_{\bm{k}}$ is the velocity operator,
and we put $G^{\mathrm{R(\bm{A})}} = G^{\mathrm{R(A)}}_{\bm{k}}(\varepsilon)$ for simplicity.
Here we assume that the spatial variation of the temperature $T(\bm{r})$ is slow and use the relation 
$(\partial_{i} f_{\mathrm{leq}}) = (\partial_{i}T/T) \left( - \partial f/\partial \varepsilon \right) (\varepsilon - \mu)$
with (global) equilibrium distribution function $f(\varepsilon)$.
The detailed derivation of the NCTE Hall current
is shown in Appendix.

Equation~(\ref{eq:NCTE_general}) is one of the main results of this letter.
By examining Eq.~(\ref{eq:NCTE_general}), 
we find that the NCTE Hall current becomes zero when the temperature reaches absolute zero, i.e., $T \to 0$.
We can also find that the non-commutativity of the velocity operators 
is a critical factor for the NCTE Hall effect,
which implies that the NCTE Hall effect occurs in the chiral materials.
We can show from Eq.~(\ref{eq:NCTE_conductivity}) that chiral is the only required symmetry for the NCTE Hall effect.
It is worth noting that, unlike the conventional (anomalous) Hall effect,
the NCTE Hall effect does not necessarily require time-reversal breaking as an essential condition.

%%%%%%%%%% -------------------------------------------------- %%%%%%%%%%

%%%%%%%%%% -------------------------------------------------- %%%%%%%%%%

\textit{NCTE Hall effect in the band representation.--}
To describe the NCTE Hall current in the band representation, 
we introduce the unitary matrix $U_{n}(\bm{k})$, which diagonalizes $H_{\bm{k}}$ such that
 $U_{n}^{\dagger}(\bm{k}) H_{\bm{k}} U_{n}(\bm{k}) = \varepsilon_{n\bm{k}}$, 
where $\varepsilon_{n\bm{k}}$ represents the eigenenergy of band $n$.
We define the retarded Green's function in band representation as
$G^{\mathrm{R}}_{n\bm{k}}(\varepsilon) = [\varepsilon - \varepsilon_{n\bm{k}} + i \gamma]^{-1}$, 
where we assume the self-energy is written as $\Sigma^{\mathrm{R}} = - i \gamma$.
We drop any dependence on energy, momentum, and band indices for the damping rate $\gamma$.
This assumption is essentially the same as in previous studies~\cite{Toshio2020,Nakai2019}
by defining the relaxation time as $\tau = 1/(2 \gamma)$.
We obtain the NCTE Hall conductivity with $\sigma^{\mathrm{NCTE}}_{z} = \sigma_z^{\rm BC} + \sigma_z^{\rm OM}$;
\begin{align}
	\sigma^{\mathrm{BC}}_{z}
	=&
	- e^{2} \tau \sum_{n,\bm{k}}
	\left( \varepsilon_{n\bm{k}} - \mu \right)
	\left( - \frac{\partial f}{\partial \varepsilon} \right)_{\varepsilon = \varepsilon_{n\bm{k}}}
	\notag \\
	& \times
	\left[ 
		(\partial_{z} \varepsilon_{n\bm{k}}) \Omega^{z}_{n}
		-
		\frac{1}{2}
		\left\{ 
			(\partial_{x} \varepsilon_{n\bm{k}}) \Omega^{x}_{n}
			+
			(\partial_{y} \varepsilon_{n\bm{k}}) \Omega^{y}_{n}
		\right\}
	\right],
	\label{eq:NCTE_BC}
	\\
	\sigma^{\mathrm{OM}}_{z}
	%=&
	%e \tau
	%\sum_{n,\bm{k}}
	%\partial_{\varepsilon}
	%\left\{ 
	%	\left( \varepsilon - \mu \right)
	%	\left( - \frac{\partial f}{\partial \varepsilon} \right)
	%\right\}_{\varepsilon = \varepsilon_{n\bm{k}}}
	%\notag \\
	%& \times
	%\left[ 
	%	(\partial_{x} \varepsilon_{n\bm{k}}) m^{x}_{n\bm{k}}
	%	+
	%	(\partial_{y} \varepsilon_{n\bm{k}}) m^{y}_{n\bm{k}}
	%\right]
	%\notag \\
	=&
	-e \tau
	\sum_{n,\bm{k}}
	\left( \varepsilon_{n\bm{k}} - \mu \right)
	\left( - \frac{\partial f}{\partial \varepsilon} \right)_{\varepsilon = \varepsilon_{n\bm{k}}}
	\bm{\nabla}_{\bm{k}} \cdot \bm{m}^{\perp}_{n\bm{k}}
	,
	\label{eq:NCTE_OM}
\end{align}
where $\bm{\Omega}_{n} = \bm{\nabla} \times \bm{A}_{n}(\bm{k})$ is the Berry curvature 
with the Berry connection $\bm{A}_{n}(\bm{k}) = - i U^{\dagger}_{n}(\bm{k}) \bm{\nabla} U_{n}(\bm{k})$,
and 
$\bm{m}_{n\bm{k}}^{\perp} = \bm{m}_{n\bm{k}} - m^{z}_{n\bm{k}} \hat{\mathrm{e}}_{z} 
= (m^{x}_{n\bm{k}}, m^{y}_{n\bm{k}}, 0)$ 
is the orbital magnetic moment
which is written by the ``interband Berry curvature'' as
\begin{align}
	\bm{m}_{n\bm{k}}
	=
	\frac{e}{2}
	\sum_{m} (\varepsilon_{m\bm{k}} - \varepsilon_{n\bm{k}})
	\Im \left[ \bm{A}_{nm}(\bm{k}) \times \bm{A}_{mn}(\bm{k}) \right],
	\label{eq:orbital_magnetic_moment}
\end{align}
where $\bm{A}_{nm}(\bm{k}) = - i U^{\dagger}_{n} \bm{\nabla} U_{m}$
is the ``interband Berry connection.'' 
Equation~(\ref{eq:orbital_magnetic_moment}) is equivalent to the expression used in Ref.~\cite{Yoda2018}.
$\sigma^{\mathrm{BC}}_{z}$ can be expressed by the Berry curvature dipole which has been %obtained based on
discussed within semiclassical theory~\cite{Nakai2019,Toshio2020}. 
On the other hand, $\sigma^{\mathrm{OM}}_{z}$ is a novel term that we have identified,  
which is comparable with the conventional term $\sigma^{\mathrm{BC}}$ %since they have the same order in
in terms of the relaxation time $\tau$.
In the following, we show that $\sigma^{\mathrm{OM}}_{z}$ is essential in a concrete model.
We note that there is no need to assume the (constant) relaxation time approximation in Eq.~(\ref{eq:NCTE_general}),
since all contributions, 
such as energy and wavenumber dependencies, can be taken into account naturally by calculating the self-energy
and corresponding vertex corrections.

%%%%%%%%%% -------------------------------------------------- %%%%%%%%%%

%%%%%%%%%% -------------------------------------------------- %%%%%%%%%%

%%%%% ----- Figure ----- %%%%%
\begin{figure}[t]
	\begin{center}
		\includegraphics[width=8.6cm]{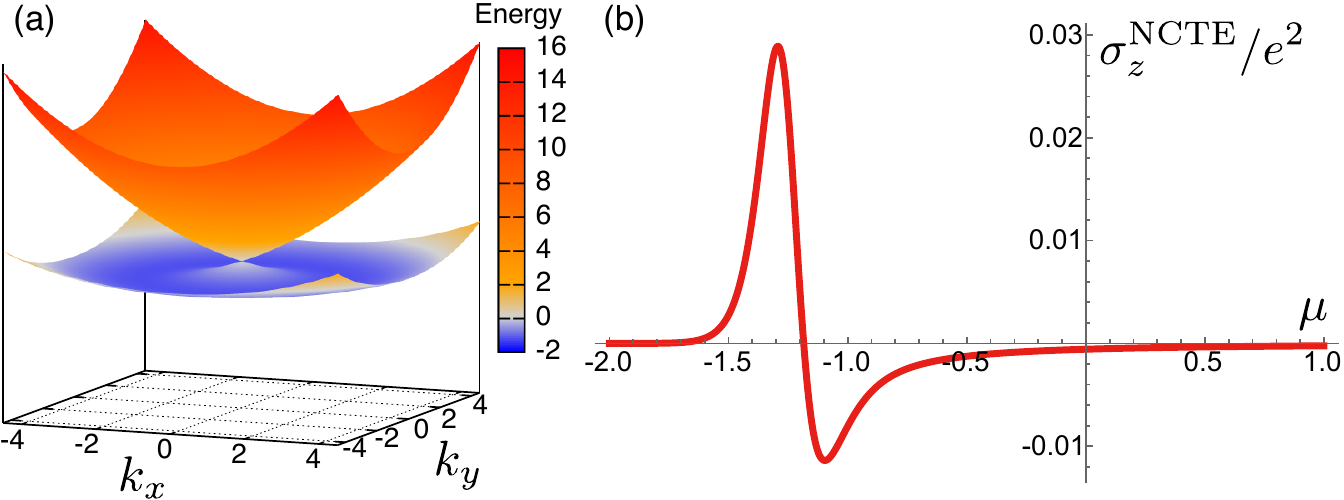}
		\caption{(a) Band structure of Hamiltonian~{(\ref{eq:Hamiltonian_minimal})} at $k_{z} = 0$.
			(b) The chemical potential dependence of the NCTE Hall conductivity at temperature $T = 0.05$.
			Here we fix the parameters as $v_{\mathrm{F}} = 1$, $\lambda = 0.2$ and $\gamma = 0.1$.
		\vspace{-3ex}}
		\label{fig:band_NCTE}
	\end{center}
\end{figure}
%%%%% ----- END Figure ----- %%%%%

\textit{Analysis in the minimal model.--}
Here we give a minimal model in which the finite NCTE Hall conductivity arises.
This model consists of the Weyl electrons (linear in wave vector) with a term of second order in wave vector, 
which is written as
\begin{align}
	H_{\bm{k}}
	=&
	v_{\mathrm{F}} \bm{k} \cdot \bm{\sigma}
	+
	\lambda k^{2},
	\label{eq:Hamiltonian_minimal}
\end{align}
where $v_{\mathrm{F}}$ is the Fermi velocity,
$\bm{\sigma} = (\sigma^{x}, \sigma^{y}, \sigma^{z})$ are the Pauli matrices,
and $\lambda$ represents the strength of a term proportional to $k^{2}$.
This model holds the time-reversal symmetry, and the anomalous Hall effect does not occur. 
The eigenenergy $\varepsilon_{\pm}$ and eigenvectors $\bm{u}_{\pm}$ are given as
$H_{\bm{k}} \bm{u}_{\pm} = \varepsilon_{\pm} \bm{u}_{\pm}$ with
\begin{align}
	\varepsilon_{\pm}
    =&
	\lambda k^{2} \pm v_{\mathrm{F}} k,
    \label{eq:eigenenergy}
    \\
    \bm{u}_{+}
    =&
	\mqty(
		\cos \frac{\theta}{2} \mathrm{e}^{- i \varphi/2} \\ 
		\sin \frac{\theta}{2} \mathrm{e}^{i \varphi/2}
	),
    \label{eq:eigenvectorsp}
	\\
    \bm{u}_{-}
    =&
	\mqty(
		\sin \frac{\theta}{2} \mathrm{e}^{- i \varphi/2} \\ 
		-\cos \frac{\theta}{2} \mathrm{e}^{i \varphi/2}
	),
    \label{eq:eigenvectorsm}
\end{align}
where we introduce the polar coordinates $(k,\theta,\varphi)$ with 
$k_{x} = k \sin \theta \cos \varphi, \ k_{y} = k \sin \theta \sin \varphi, \ k_{z} = k \cos \theta$.
The energy band structure at $k_{z} = 0$ is shown in Fig.~\ref{fig:band_NCTE}~(a).
The eigenvectors in this model are the same as in the Weyl model, 
and we can calculate the Berry curvature and the orbital magnetic moment, 
\begin{align}
	\bm{\Omega}_{\pm} 
	=&
	\pm \frac{\bm{k}}{k^{3}},
    \label{eq:Berrycurvature}
    \\
	\bm{m}_{\pm}
    =&
	- \frac{e v_{\mathrm{F}}}{2} \frac{\bm{k}}{k^{2}}.
    \label{eq:orbital_magnetic_moment_simple}
\end{align}
Substituting (\ref{eq:Berrycurvature}), (\ref{eq:orbital_magnetic_moment_simple})
and the velocity $v_{\pm,j} = \partial_{k_{j}} \varepsilon_{\pm} = \left( 2 \lambda \pm v_{\mathrm{F}}/k \right) k_{j}$
in (\ref{eq:NCTE_BC}) and (\ref{eq:NCTE_OM}), 
we obtain the NCTE Hall conductivities
\begin{align}
    \sigma^{\mathrm{BC}}_{z} 
	=& 0,
    \label{eq:sigma_BC_simple}
    \\
    \sigma^{\mathrm{OM}}_{z} 
	=&
	\frac{e^{2}v_{\mathrm{F}} \tau}{3 \pi^{2}}
	\int_{-\frac{v_{\mathrm{F}}^{2}}{4 \lambda}}^{\infty}
	\mathrm{d} \varepsilon
	\frac{\varepsilon - \mu}{\sqrt{v_{\mathrm{F}}^{2} + 4 \lambda \varepsilon}}
	\left( - \frac{\partial f}{\partial \varepsilon} \right).
    \label{eq:sigma_OM_simple}
\end{align}
We find that the contribution from the Berry curvature dipole disappears, 
while the contribution from the orbital magnetic moment is essential. 
The dependence of the NCTE Hall conductivity on the chemical potential is shown in Fig.~\ref{fig:band_NCTE}~(b). 
Although the orbital magnetic moment arises near the "Weyl point" $\varepsilon_{\bm{k}}=0$,
the NCTE Hall conductivity is enhanced when we tune the chemical potential near the bottom of the lower energy band. 
This enhancement reflects the nature of thermoelectric transport,
which is enhanced when the chemical potential is near a sharp singularity in the density of states~\cite{Mahan1996}. 
We can also see that the NCTE Hall effect is zero in the linear model, see Appendix.
We emphasize that we need to consider not only the enhancement of the Berry curvature dipoles or the orbital magnetic moment 
but also the differentials of the density of states to obtain the large NCTE Hall conductivity.

%%%%%%%%%% -------------------------------------------------- %%%%%%%%%%

%%%%%%%%%% -------------------------------------------------- %%%%%%%%%%

%%%%% ----- Figure ----- %%%%%
\begin{figure}[t]
	\begin{center}
		\includegraphics[width=8.6cm]{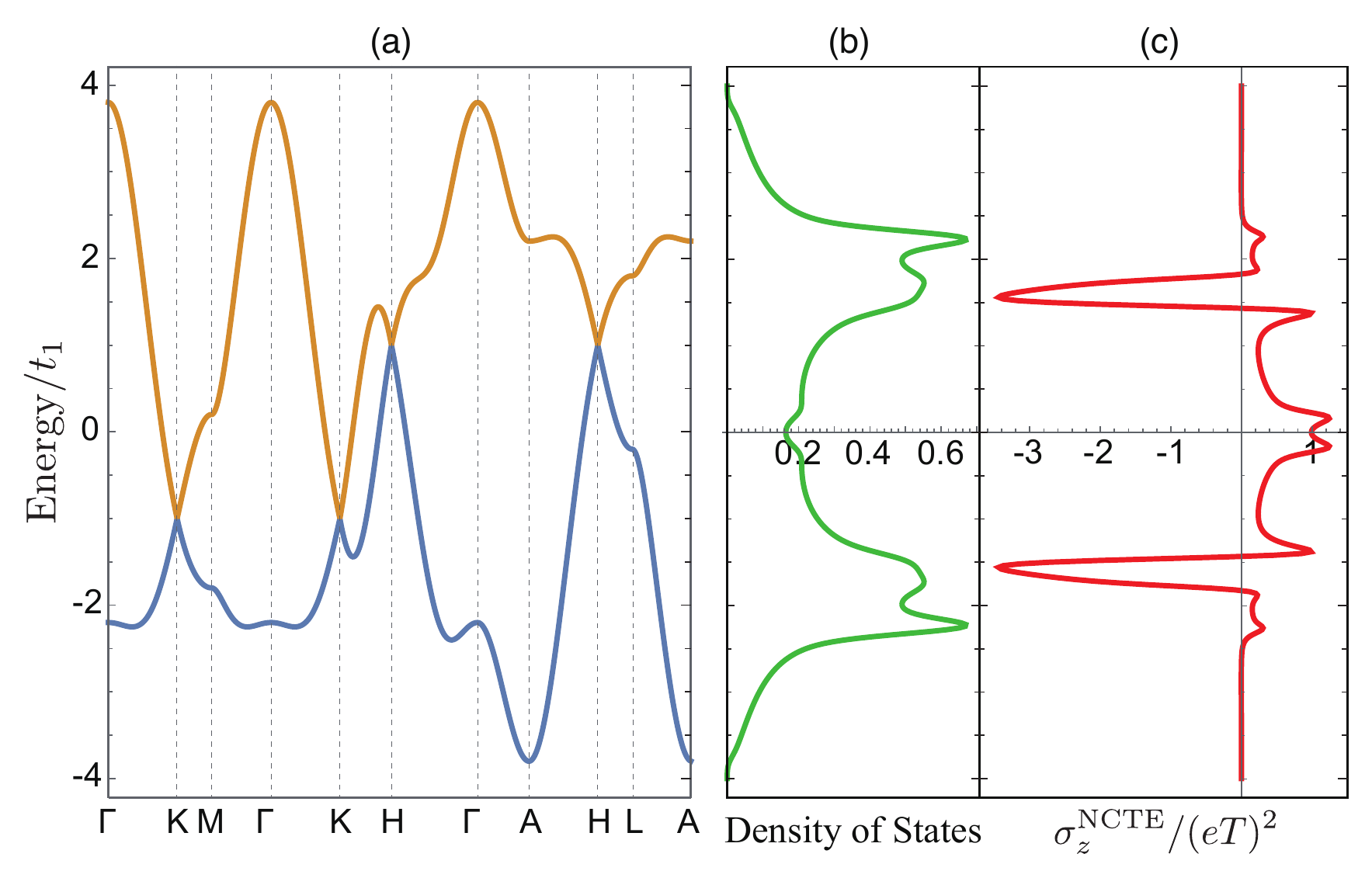}
		\caption{(a) Energy bands and (b) density of states of the Hamiltonian~(\ref{eq:Hamiltonian_Yoda}).
		(c) The NCTE Hall conductivity within the lowest order in Sommerfeld expantion as a function of the chemical potential.
		We use the parameters $a=1.0, c=1.0, t_1 = 1.0, t_2 = 0.2, t_3 = -0.2$ and $\gamma = 0.1$.
		\vspace{-3ex}}
		\label{fig:Yoda_NCTE}
	\end{center}
\end{figure}
%%%%% ----- END Figure ----- %%%%%

\begin{figure}[t]
	\begin{center}
		\includegraphics[width=8.6cm]{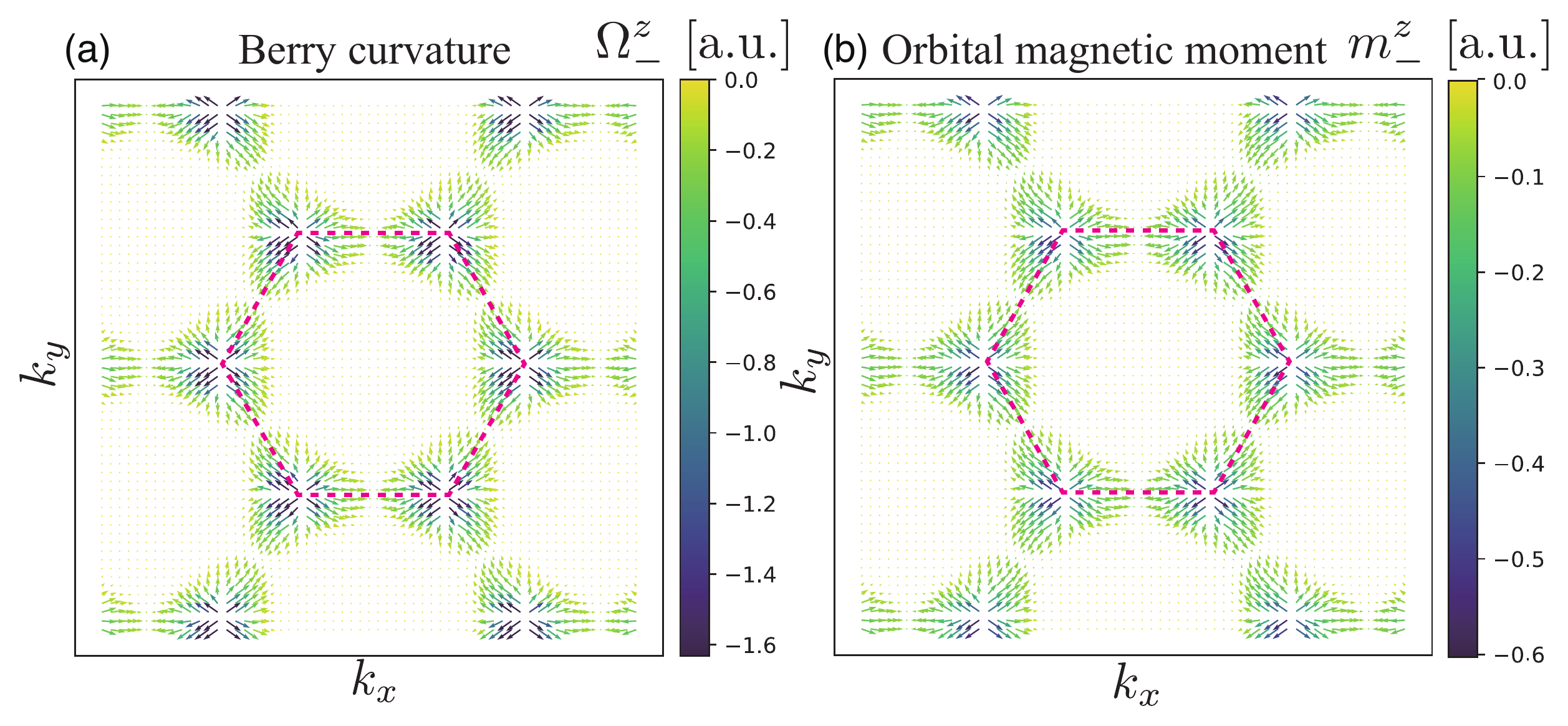}
		\caption{(a) The Berry curvature and (b) the orbital magnetic moment of the lower band at $k_{z} = \pi - \delta$,
		where we put the small quantity $\delta = 3 \pi/160$ to avoid the singular points.
		The length of arrows represent the magnitude of them in $k_x$-$k_y$ plane,
		and color represents the magnitude of them in $k_z$ direction.
		Dashed lines represent the boundary of the Brillouin zones.
		Both the Berry curvature and the orbital magnetic moment are enhanced near the H points.
		\vspace{-3ex}}
		\label{fig:Berry_OM}
	\end{center}
\end{figure}

\textit{NCTE Hall effect in chiral crystal model.--}
Theoretical studies predict that chiral crystals exhibit nontrivial transport phenomena~\cite{Yoda2015,Yoda2018}
and realize in trigonal Te and Se~\cite{Hirayama2015-bv}. 
Here we demonstrate the NCTE Hall effect in a model of chiral Weyl semimetal\cite{Armitage2018-qr}. 
We consider the tight-binding model proposed in references~\cite{Yoda2015,Yoda2018}, 
which consists of an infinite stack of honeycomb lattice layers, and describe the Bloch Hamiltonian as 
\begin{align}
	H_{\bm{k}} 
	=&
	d_{0} + \bm{d}_{\bm{k}} \cdot \bm{\sigma},
	\label{eq:Hamiltonian_Yoda}
	\\
	d_{0}
	=&
	2 t_{2} \cos (k_{z} c) \sum_{i} \cos (\bm{k} \cdot \bm{b}_{i})
	+ 2 t_{3} \cos (k_{z} c),
	\\
	d_{x}
	=&
	t_{1} \sum_{i} \cos (\bm{k} \cdot \bm{a}_{i}),
	\\
	d_{y}
	=&
	t_{1} \sum_{i} \sin (\bm{k} \cdot \bm{a}_{i}),
	\\
	d_{z}
	=&
	- 2 t_{2} \sin (k_{z} c) \sum_{i} \sin (\bm{k} \cdot \bm{b}_{i}),
\end{align}
where $\bm{d}_{\bm{k}} = (d_{x}, d_{y}, d_{z})$,
$\bm{\sigma} = (\sigma^{x}, \sigma^{y}, \sigma^{z})$ are the Pauli matrices,
$\bm{a}_{1} = a/\sqrt{3} \hat{\mathrm{e}}_{y}$,
$\bm{a}_{2} = a/2 \hat{\mathrm{e}}_{x} - a/(2 \sqrt{3}) \hat{\mathrm{e}}_{y}$,
$\bm{a}_{3} = - a/2 \hat{\mathrm{e}}_{x} - a/(2 \sqrt{3}) \hat{\mathrm{e}}_{y}$,
$\bm{b}_{1} = a \hat{\mathrm{e}}_{x}$,
$\bm{b}_{2} = - a/2 \hat{\mathrm{e}}_{x} + a \sqrt{3}/2 \hat{\mathrm{e}}_{y}$,
$\bm{b}_{3} = - a/2 \hat{\mathrm{e}}_{x} - a \sqrt{3}/2 \hat{\mathrm{e}}_{y}$, 
$t_{1}, t_{2}, t_{3}$ are the hopping parameters, 
and $a$ and $c$ are the intralayer and interlayer lattice constants, respectively.
We plot the energy bands and the density of states of the Hamiltonian~(\ref{eq:Hamiltonian_Yoda}) 
in Fig.~\ref{fig:Yoda_NCTE}~(a) and (b), respectively. 
(Parameters we use in the calculation are shown in the caption of Fig.~\ref{fig:Yoda_NCTE}.) 
There are Weyl points at K and H points, and the Berry curvature and the orbital magnetic moment are enhanced around the Weyl nodes, 
as shown in Fig.~\ref{fig:Berry_OM}. 
To obtain the NCTE Hall conductivity, 
we apply the Sommerfeld expansion 
and evaluate the lowest order in temperature $T$, assuming a low-temperature limit. 
We also assume a constant and pure imaginary self-energy $\Sigma^{\mathrm{R}} = - i \gamma$. 
In Fig.~\ref{fig:Yoda_NCTE}~(c),
we plot the NCTE Hall conductivity of this model at a low temperature as a function of the chemical potential.
We find a finite NCTE Hall conductivity in a wide range of the chemical potential, 
with an enhancement of the NCTE Hall conductivity near the Weyl points 
and at points where the density of states varies sharply, 
which is consistent with the discussion in the minimal model. 
We estimate the NCTE Hall conductivity in this model using realistic parameters. 
In Fig.~\ref{fig:Yoda_NCTE} (c), we set $\gamma/t_{1} = 0.1$ for numerical calculation, which is too large for a realistic situation. 
As given in eq.(\ref{eq:NCTE_general}), 
we assume that the NCTE Hall conductivity is proportional to $\tau = \hbar/(2 \gamma)$ and set $\gamma/t_{1} = 0.001$ for estimation.
Assuming a temperature gradient $\nabla T /T \simeq 10^{2}$~m$^{-1}$, 
an electric field $E \simeq 10^{3}$~V/m,
and a system size $L \times L \times L$ with $L \simeq 1.0$~$\mu$m,
we obtain an NCTE Hall conductivity of order $\sigma^{\mathrm{NCTE}} \sim 10 e^2/h$
and corresponding NCTE Hall current of order $j_{z} \sim 100$~pA, which are experimentally measurable.

%%%%% ----- Figure ----- %%%%%
%%%%% ----- END Figure ----- %%%%%

%%%%%%%%%% -------------------------------------------------- %%%%%%%%%%

%%%%%%%%%% -------------------------------------------------- %%%%%%%%%%

\textit{Conclusion.--}
In this letter, we have formulated the NCTE Hall effect using the method of quantum field theory at the microscopic level. 
By rewriting the formula in the band representation, we have identified the contributions of the orbital magnetic moment 
in addition to the Berry curvature dipole contributions. 
Through analysis of the minimal model, 
we have demonstrated the essential nature of the contribution of the orbital magnetic moment to the NCTE Hall effect. 
Our findings show that the chiral crystal model gives rise to finite NCTE Hall conductivity. 
These results provide important insights into the understanding of nonlinear transport phenomena 
and pave the way for further investigations in this field.

%%%%%%%%%% -------------------------------------------------- %%%%%%%%%%

\begin{acknowledgments}
 We acknowledge G. Qu and E. Saitoh for valuable discussion.
 AY is supported by
 JSPS KAKENHI (Grant No.~JP20K03835) and the Sumitomo Foundation (190228).
 KN is supported by JSPS KAKENHI (Grant No. JP21K13875).
 TY is supported by JSPS KAKENHI (Grant No.~JP21K14526).
\end{acknowledgments}

\appendix

\onecolumngrid

\section{Derivation of the nonlinear chiral thermo-electric Hall current}

In this section, we derive the formula of nonlinear chiral thermo-electric (NCTE) Hall current
based on the method of quantum field theory with the nonequilibrium (Keldysh) Green's function.

\subsection{Brief introduction of Green's function in Keldysh space}

The expectation value of physical quantity $\hat{O}$ can be calculate by using the nonequilibrium (Keldysh) Green's function as
\begin{align}
	\langle \hat{O} \rangle
	=
	- \frac{i}{2} \mathsf{Tr} \left[ (\hat{O} \otimes \gamma^{\mathrm{q}}) \check{G} \right]
	=
	- \frac{i}{2} \mathrm{tr} \left[ \hat{O} G^{\mathrm{K}} \right],
	\label{eq:ave_O}
\end{align}
where $\check{G}$ is the Green's function (matrix) in Keldysh space,
\begin{align}
	\check{G}
	=
    \pmqty{
		G^{\mathrm{R}} & G^{\mathrm{K}} \\
		0 & G^{\mathrm{A}}
	},
	\label{eq:G_Kel}
\end{align}
and we define a matrix $\gamma^{\mathrm{q}}$ in Keldysh space as
\begin{align}
	\gamma^{\mathrm{q}} = \pmqty{0 & 1 \\ 1 & 0}.
\end{align}
The symbol $\mathrm{tr}[\cdots]$ means the trace and sum (integral) in all quantum numbers,
and $\mathsf{Tr}[\cdots]$ includes the trace in Keldysh space in addition to $\mathrm{tr}[\cdots]$.
Equation (\ref{eq:ave_O}) means that we can obtain the expectation value of any physical quantity 
if we get $G^{\mathrm{K}}$.

When the unperturbed Hamiltonian has time- and space-translational symmetries,
the system can be described as the action
\begin{align}
	S_{0}
	=
	\int \frac{\mathrm{d} \varepsilon}{2 \pi} \sum_{\bm{k}}
	\bar{\psi}^{\alpha}_{n\bm{k}}(\varepsilon)
	\left[ g^{-1}_{nm}(\bm{k},\varepsilon) \right]^{\alpha \beta}
	\psi^{\beta}_{m\bm{k}}(\varepsilon),
	\label{eq:S0}
\end{align}
where $\alpha, \beta$ represent the indices in Keldysh space.
Then the unperturbed Green's function is expressed as
\begin{align}
	g^{\alpha \beta}_{nm}(\bm{k},\varepsilon)
	=
	- i \langle \psi^{\alpha}_{n\bm{k}}(\varepsilon) \bar{\psi}^{\beta}_{m\bm{k}}(\varepsilon) \rangle_{S_{0}}
	=
	- i \int \mathcal{D} \left[ \bar{\psi}, \psi \right]
	\psi^{\alpha}_{n\bm{k}}(\varepsilon) \bar{\psi}^{\beta}_{m\bm{k}}(\varepsilon)
	\mathrm{e}^{i S_{0}}.
\end{align}
Here we consider the external field given as
\begin{align}
	S_{\mathrm{ext}}
	=
	-
	\int \frac{\mathrm{d}\omega}{2 \pi} \sum_{\bm{q}}
	\hat{A}_{-\bm{q}}(- \omega) \Phi_{\bm{q}}(\omega),
	\label{eq:Sext}
\end{align}
where 
$\hat{A}_{-\bm{q}}(-\omega) 
= 
\int (\mathrm{d}\varepsilon/(2\pi)) \sum_{\bm{k}} \sum_{\gamma}
\bar{\psi}^{\gamma}_{n,\bm{k}+\bm{q}/2} (\varepsilon + \omega/2)
A_{nm}(-\bm{q},-\omega) \psi^{\gamma}_{m,\bm{k}-\bm{q}/2}(\varepsilon - \omega/2)$ 
is the physical (quantum) quantity coupled to the external (classical) field $\Phi_{\bm{q}}(\omega)$.
The Green's function with the external field $G(\bm{k},\bm{k}'; \varepsilon, \varepsilon')$ is expressed as
\begin{align}
	G^{\alpha \beta}_{nm}(\bm{k},\bm{k}'; \varepsilon, \varepsilon')
	=
	- i \langle \psi^{\alpha}_{n\bm{k}}(\varepsilon) \bar{\psi}^{\beta}_{m\bm{k}'}(\varepsilon') \rangle_{S_{0}+S_{\mathrm{ext}}}
	=
	- i \int \mathcal{D} \left[ \bar{\psi}, \psi \right]
	\psi^{\alpha}_{n\bm{k}}(\varepsilon) \bar{\psi}^{\beta}_{m\bm{k}'}(\varepsilon')
	\mathrm{e}^{i S_{0} + S_{\mathrm{ext}}}.
	\label{eq:G_tot}
\end{align}
Expanding Eq.(\ref{eq:G_tot}) in the external field $\Phi$, 
we obtain the Green's function with any order in $\Phi$.
For example, the Green's function with the first order in $\Phi$, denotes $G_{1}$, are expressed as
\begin{align}
	G^{\alpha \beta}_{1,nm} (\bm{k}, \varepsilon ; \bm{q}, \omega)
	=&
	g^{\alpha \gamma}_{n \ell_{1}} \left( \bm{k} + \frac{\bm{q}}{2}, \varepsilon + \frac{\omega}{2} \right)
	A_{\ell_{1}, \ell_{2}}(-\bm{q}, - \omega)
	g^{\gamma \beta}_{\ell_{2} m} \left( \bm{k} - \frac{\bm{q}}{2}, \varepsilon - \frac{\omega}{2} \right)
	\Phi_{\bm{q}}(\omega)
	\notag \\
	=&
	\left[ 
		\check{g} \left( \bm{k} + \frac{\bm{q}}{2}, \varepsilon + \frac{\omega}{2} \right)
		A(-\bm{q}, - \omega)
		\check{g} \left( \bm{k} - \frac{\bm{q}}{2}, \varepsilon - \frac{\omega}{2} \right)
	\right]^{\alpha \beta}_{nm}
	\Phi_{\bm{q}}(\omega),
	\label{eq:G_linear}
\end{align}
and focusing on the Keldysh component, we obtain
\begin{align}
	G^{\mathrm{K}}_{1}(\bm{k}, \varepsilon ; \bm{q}, \omega)
	=&
	\left[ 
		g^{\mathrm{K}}_{\bm{k}+\frac{\bm{q}}{2}}\left( \varepsilon + \frac{\omega}{2} \right)
		A(-\bm{q}, - \omega)
		g^{\mathrm{A}}_{\bm{k}-\frac{\bm{q}}{2}}\left( \varepsilon - \frac{\bm{q}}{2} \right)
	\right.
	\notag \\
	& 
	\left.
		+
		g^{\mathrm{R}}_{\bm{k}+\frac{\bm{q}}{2}}\left( \varepsilon + \frac{\omega}{2} \right)
		A(-\bm{q}, - \omega)
		g^{\mathrm{K}}_{\bm{k}-\frac{\bm{q}}{2}}\left( \varepsilon - \frac{\bm{q}}{2} \right)
	\right]
	\Phi_{\bm{q}}(\omega).
	\label{eq:GK_linear}
\end{align}
This expression (\ref{eq:GK_linear}) is consistent with the consequence of the Langrerh method.
Assuming that the system reaches thermal equilibrium before applying the external field, 
the Keldysh component of the unperturbed Green's function is given as
\begin{align}
	g^{\mathrm{K}}_{\bm{k}}(\varepsilon)
	=
	\left[ 1 - 2 f(\varepsilon) \right]
	\left( g^{\mathrm{R}}_{\bm{k}}(\varepsilon) - g^{\mathrm{A}}_{\bm{k}}(\varepsilon) \right),
	\quad
	f(\varepsilon)
	=
	\frac{1}{\mathrm{e}^{(\varepsilon - \mu)/T} + 1},
	\label{eq:gK_unp}
\end{align}
with temperature $T$ and chemical potential $\mu$, 
then we obtain the expectation value of physical quantity $O$ by using Eq.(\ref{eq:GK_linear}) as
\begin{align}
	\langle O \rangle_{\Phi^{1}} (\bm{q},\omega)
	=&
	i
	\int \frac{\mathrm{d}\varepsilon}{2 \pi} \sum_{\bm{k}}
	\mathrm{tr}
	\left[ 
		\left( f(\varepsilon_{+}) - f(\varepsilon_{-}) \right)
		O(\bm{q},\omega)
		g^{\mathrm{R}}_{\bm{k}_{+}}(\varepsilon_{+})
		A(-\bm{q}, - \omega)
		g^{\mathrm{A}}_{\bm{k}_{-}}(\varepsilon_{-})
	\right.
	\notag \\
	& 
	\left.
		+
		f(\varepsilon_{-}) 
		O(\bm{q},\omega)
		g^{\mathrm{R}}_{\bm{k}_{+}} (\varepsilon_{+}) 
		A(-\bm{q},-\omega) 
		g^{\mathrm{R}}_{\bm{k}_{-}} (\varepsilon_{-})
	\right.
	\notag \\
	& 
	\left.
		-
		f(\varepsilon_{+}) 
		O(\bm{q},\omega)
		g^{\mathrm{A}}_{\bm{k}_{+}} (\varepsilon_{+}) 
		A(-\bm{q},-\omega) 
		g^{\mathrm{A}}_{\bm{k}_{-}} (\varepsilon_{-})
	\right]
	\Phi_{\bm{q}}(\omega),
	\label{eq:Kubo}
\end{align}
where we define $\bm{k}_{\pm} = \bm{k} \pm \bm{q}/2$ and $\varepsilon_{\pm} = \varepsilon \pm \omega/2$
for simplicity.
This can be obtained starting from the Kubo formula.

\subsection{Response to the gradient of external potential}
In main text, we consider the (electromagnetic) scalar potential which couples the electron density,
\begin{align}
	S_{\mathrm{ext}}
	=
	- e \int \bm{d} t \int \mathrm{d} \bm{r} \ \hat{n}(\bm{r},t) \phi(\bm{r},t),
	\label{eq:Hext}
\end{align}
and we want to calculate the response to the electric field $\bm{E} = - \bm{\nabla} \phi$.
To obtain this, we consider the continuous equation
\begin{align}
	\frac{\partial \hat{n}}{\partial t}
	+
	\bm{\nabla} \cdot \hat{\bm{j}}
	= 0.
	\label{eq:continuous_equation}
\end{align}
Now we write Eq. (\ref{eq:continuous_equation}) in Fourier space.
Using the relations
\begin{align}
	- i \omega \hat{n}_{\bm{q}}(\omega)
	=&
	(\partial_{t} \hat{n})_{\omega} - (\partial_{t} \hat{n})_{\omega=0},
	\\
	i \bm{q} \hat{n}_{\bm{q}}(\omega)
	=&
	(\bm{\nabla} \hat{n})_{\bm{q}},
\end{align}
we can rewrite the action as
\begin{align}
	S_{\mathrm{ext}}
	=&
	- e
	\int \frac{\mathrm{d}\omega}{2 \pi}
	\sum_{\bm{q}}
	\hat{n}_{-\bm{q}}(-\omega) \phi_{\bm{q}}(\omega)
	\notag\\
	=&
	- e
	\int \frac{\mathrm{d}\omega}{2 \pi}
	\sum_{\bm{q}}
	\frac{i \omega \hat{n}_{-\bm{q}}(-\omega)}{i \omega} \phi_{\bm{q}}(\omega)
	\notag \\
	=&
	- e
	\int \frac{\mathrm{d}\omega}{2 \pi}
	\sum_{\bm{q}}
	\frac{1}{i \omega}
	\left[ 
		(\partial_{t} \hat{n})_{-\omega} - (\partial_{t} \hat{n})_{0}
	\right]
	\phi_{\bm{q}}(\omega)
	\notag \\
	=&
	- e
	\int \frac{\mathrm{d}\omega}{2 \pi}
	\sum_{\bm{q}}
	\frac{i q_{i}}{i \omega}
	\left[ 
		j_{i,-\bm{q}}(-\omega) - j_{i,-\bm{q}}(0)
	\right]
	\phi_{\bm{q}}(\omega)
	\notag \\
	=&
	e
	\int \frac{\mathrm{d}\omega}{2 \pi}
	\sum_{\bm{q}}
	\frac{1}{i \omega}
	\left[ 
		j_{i,-\bm{q}}(-\omega) - j_{i,-\bm{q}}(0)
	\right]
	(- \partial_{i} \phi)_{\bm{q}}(\omega).
\end{align}
Therefore, we have to subtract the $\omega = 0$ component from finite $\omega$ calculation
when we calculate the gradient of the external field.
Note that $\bm{j} = \bm{v}_{\bm{k}} = \bm{\nabla}_{\bm{k}} H_{\bm{k}}$ is satisfied in our situation.

\subsection{Response to temperature gradient}
Here, we consider the response to the temperature gradient.
To expand the general order in driving forces, 
we have to formulate the response to the temperature gradient without relying 
the relations established in the linear response theory (i.e., the Einstein-Luttinger relations).
As the same manner in the Boltzmann theory of nonlinear heat responses,
we start the assumption of the local equilibrium.
In the Boltzmann theory, the local equilibrium distribution function are introduced as the ``initial state''
and incorporate the nonequilibrium natures by solving the Boltzmann equation.
In terms of the method of nonequilibrium Green's function,
the nonequilibrium natures are incorporated by solving the Dyson's equation in Keldysh space, instead of the Boltzmann equation.
The Dyson's equation of Keldysh component is given as
\begin{align}
	G^{\mathrm{K}}
	=
	G^{\mathrm{R}} \star \Sigma^{\mathrm{K}} \star G^{\mathrm{A}},
	\label{eq:Dyson}
\end{align}
where the product $\star$ represents the convolution integral in real time-space representation,
or the Moyal product in Wigner representation.
The Keldysh component of the self-energy is written with the (local) equilibrium distribution function as
\begin{align}
	\Sigma^{\mathrm{K}}
	=&
	\left[ 1 - 2 f(\varepsilon;\bm{r}) \right] \left( \Sigma^{\mathrm{R}} - \Sigma^{\mathrm{A}} \right),
	\label{eq:selfenergy_K}
	\\
	f(\varepsilon; \bm{r})
	=&
	\frac{1}{\mathrm{e}^{(\varepsilon - \mu)/T(\bm{r})} + 1},
\end{align}
where $\Sigma^{\mathrm{R(A)}}$ is the retarded (advanced) component of self-energy.
Here we consider the Fourier component of local equilibrium distribution function,
$f(\varepsilon,\bm{r}) = f_{\bm{q}} \mathrm{e}^{i \bm{q}\cdot \bm{r}}$,
and using the relation $\Sigma^{\mathrm{R(A)}} = \varepsilon - H_{\bm{k}} - (G^{\mathrm{R(A)}}_{\bm{k}}(\varepsilon))$,
we rewrite the Dyson's equation (\ref{eq:Dyson}) as
\begin{align}
	G^{\mathrm{K}}_{\bm{k},\bm{q}}(\varepsilon)
	=&
	(1 - 2 f_{\bm{q}}(\varepsilon))
	\left[ G^{\mathrm{R}}_{\bm{k}+\frac{\bm{q}}{2}}(\varepsilon)- G^{\mathrm{A}}_{\bm{k}-\frac{\bm{q}}{2}} (\varepsilon) \right]
	\notag \\
	& -
	(1 - 2 f_{\bm{q}}(\varepsilon))
	G^{\mathrm{R}}_{\bm{k}+\frac{\bm{q}}{2}}(\varepsilon)
	\left[ 
		H_{\bm{k}+\frac{\bm{q}}{2}} - H_{\bm{k}- \frac{\bm{q}}{2}}
	\right]
	G^{\mathrm{A}}_{\bm{k}-\frac{\bm{q}}{2}}(\varepsilon).
	\label{eq:GK_leq}
\end{align}
The first term of Eq. (\ref{eq:GK_leq}) represents the (local) equilibrium component,
whereas the second term includes the nonequilibrium natures.
By expanding Eq. (\ref{eq:GK_leq}) in $\bm{q}$, 
which corresponds to the gradient expansion in Wigner representation,
we can obtain the Keldysh Green's function including the temperature gradient.
We note that this method can apply when we consider the response to the gradient of the chemical potential $\bm{\nabla} \mu$.

\subsection{Response to an electric field and a temperature gradient}
We formulate the nonlinear response to an electric field and a temperature gradient.
Here we consider the uniform and static electric field and ignore the wave vector dependences of electric field
and taking $\omega \to 0$.
The expectation value of the physical quantity $\mathcal{O}$ for the response to $E_{i} \partial_{j} T$
is expressed as
\begin{align}
	\langle \mathcal{O} \rangle_{E_{i} \partial_{j} T}
	=
	\lim_{\omega \to 0}
	\frac{K_{i}(\omega) - K_{i}(0)}{i \omega} E_{i},
	\label{eq:O_K}
\end{align}
where 
\begin{align}
	K_{i}(\omega)
	=
	\frac{i}{2}
	\int \frac{\mathrm{d}\varepsilon}{2 \pi}
	\sum_{\bm{k}}
	\mathrm{tr}
	\left[ 
		\mathcal{O}_{\bm{q}} 
		G^{\mathrm{K}}_{\bm{k},\bm{q}}(\varepsilon_{+})
		v_{\bm{k}_{-},i}
		G^{\mathrm{A}}_{\bm{k}_{-}}(\varepsilon_{-})
		+
		\mathcal{O}_{\bm{q}} 
		G^{\mathrm{R}}_{\bm{k}_{+}}(\varepsilon_{+})
		v_{\bm{k}_{-},i}
		G^{\mathrm{K}}_{\bm{k},\bm{q}}(\varepsilon_{-})
	\right].
	\label{eq:Ki}
\end{align}
To obtain the response to $E_{i} \partial_{j} T$,
we substitute Eq. (\ref{eq:GK_leq}) in Eq. (\ref{eq:Ki}),
expand in $\bm{q}$ and $\omega$,
and pick up the first order in $\bm{q}$ and $\omega$.
Moreover, to obtain the NCTE response,
we focus on the ``anti-symmetric'' component
$[(E_{i} \partial_{j} T) - E_{j} (\partial_{i} T)]/2$,
then we obtain
\begin{align}
	\langle \mathcal{O} \rangle_{\mathrm{NCTE}}
	=&
	\frac{1}{4 \pi}
	\int \mathrm{d} \varepsilon
	\sum_{\bm{k}}
	\left[ E_{i} (\partial_{j} f(\varepsilon))_{\bm{q}} - E_{j} (\partial_{i} f(\varepsilon))_{\bm{q}} \right]
	\notag \\
	& 
	\times
	\mathrm{Im}
	\left[ 
		\mathrm{tr}
		\left[ 
			\mathcal{O}
			\left\{ 
				g^{\mathrm{R}} v_{i} (\partial_{\varepsilon} g^{\mathrm{R}}) v_{j} g^{\mathrm{R}}
				+
				(\partial_{\varepsilon} g^{\mathrm{R}}) v_{i} g^{\mathrm{R}} v_{j} g^{\mathrm{A}}
			\right.
		\right.
	\right.
	\notag \\
	& \quad
	\left.
		\left.
			\left.
				-
				g^{\mathrm{R}} v_{i} (\partial_{\varepsilon} g^{\mathrm{R}}) v_{j} g^{\mathrm{A}}
				-
				g^{\mathrm{R}} v_{i} g^{\mathrm{R}} v_{j} (\partial_{\varepsilon} g^{\mathrm{A}})
			\right\}
		\right]
	\right].
	\label{eq:O_NCTE}
\end{align}
Using the relation
$(\partial_{i} f) = (\partial_{i} T/T) (\varepsilon - \mu) (- \partial f/\partial \varepsilon)$
and applying Eq. (\ref{eq:O_NCTE}) to the current operator, $\mathcal{O} = v_{z}$,
we obtain the formula of NCTE Hall current, shown in main text.

%%%%%%%%%% -------------------------------------------------- %%%%%%%%%%

%%%%%%%%%% -------------------------------------------------- %%%%%%%%%%

\section{Zero NCTE Hall conductivity in (anisotropic) Weyl electron system}

In this section, we show the calculation of the NCTE Hall conductivity in the (anisotropic) Weyl electron system
within constant relaxation time approximation,
and prove that the NCTE Hall conductivity should be zero.

The Hamiltonian is given as
\begin{align}
	H_{\bm{k}}
	=
	v_{\mathrm{F}} (k_{x} \sigma^{x} + k_{y} \sigma^{y})
	+
	\tilde{v}_{\mathrm{F}} k_{z} \sigma^{z}
	=
	v_{\mathrm{F}}
	(k_{x} \sigma^{x} + k_{y} \sigma^{y} + r k_{z} \sigma^{z}),
	\label{eq:Ham_aniWeyl}
\end{align}
where $v_{\mathrm{F}}$ and $\tilde{v}_{\mathrm{F}}$ are the Fermi velocities
in $x$-$y$ plane and $z$ direction, respectively,
and $r = \tilde{v}_{\mathrm{F}} / v_{\mathrm{F}}$.
The eigenvalues and eigenvectors are given as
\begin{align}
	\varepsilon_{\pm}
	=&
	\pm
	v_{\mathrm{F}}
	\sqrt{k_{x}^{2} + k_{y}^{2} + r^{2} k_{z}^{2}},
	\label{eq:eigenvalues}
	\\
	\bm{u}_{+}
	=&
	\frac{1}{\sqrt{2 \xi (\xi - r \cos \theta)}}
    \pmqty{
		\sin \theta \mathrm{e}^{- i \varphi/2}
 		\\
		(\xi - r \cos \theta) \mathrm{e}^{i \varphi /2}
	}
	=
	C
    \pmqty{
		\sin \theta \mathrm{e}^{- i \varphi/2}
 		\\
		(\xi - r \cos \theta) \mathrm{e}^{i \varphi /2}
	}
	,
	\label{eq:eigenvec_p}
	\\
	\bm{u}_{-}
	=&
	\frac{1}{\sqrt{2 \xi (\xi - r \cos \theta)}}
    \pmqty{
		(\xi - r \cos \theta) \mathrm{e}^{i \varphi /2}
 		\\
		- \sin \theta \mathrm{e}^{i \varphi/2}
	}
	=
	C
    \pmqty{
		(\xi - r \cos \theta) \mathrm{e}^{i \varphi /2}
 		\\
		- \sin \theta \mathrm{e}^{i \varphi/2}
	}
	,
	\label{eq:eigenvec_m}
\end{align}
where we put $k_{x} = k \sin \theta \cos \varphi, k_{y} = k \sin \theta \sin \varphi, k_{z} = k \cos \theta$,
$\xi = \sqrt{\sin^{2} \theta + r^{2} \cos^{2} \theta}$,
and we choose the phase factor to be $C = 1/\sqrt{2 \xi (\xi - r \cos \theta)} \in \mathbb{R}$.

From the eigenvectors (\ref{eq:eigenvec_p}) and (\ref{eq:eigenvec_m}),
we can construct the Berry connection,
\begin{align}
	A^{\sigma \sigma'}_{j}
	=
	- i \bm{u}_{\sigma}^{\dagger} \partial_{j} \bm{u}_{\sigma'},
	\label{eq:Berry_curvature}
\end{align}
where $\sigma, \sigma' = \pm$.
Using the relation $\bm{u}_{\sigma} \cdot \bm{u}_{\sigma'} = \delta_{\sigma,\sigma'}$,
we obtain 
\begin{align}
	A^{+ +}_{j}
	=&
	- (\partial_{j} \varphi) \frac{r \cos \theta}{2 \xi},
	\label{eq:App}
	\\
	A^{- -}_{j}
	=&
	(\partial_{j} \varphi) \frac{r \cos \theta}{2 \xi},
	\label{eq:Amm}
\end{align}
and
\begin{align}
	A^{+ -}_{j}
	=&
	- i C^{2} (\partial_{j} \theta)
	\left[ 
		\frac{1 - r^{2}}{\xi} \sin^{2} \theta \cos \theta + r - \xi \cos \theta
	\right]
	-
	C^{2} (\partial_{j} \varphi) \sin \theta (\xi - r \cos \theta),
	\label{eq:Apm}
	\\
	A^{-+}_{j}
	=&
	i C^{2} (\partial_{j} \theta)
	\left[ 
		\frac{1 - r^{2}}{\xi} \sin^{2} \theta \cos \theta + r - \xi \cos \theta
	\right]
	-
	C^{2} (\partial_{j} \varphi) \sin \theta (\xi - r \cos \theta).
	\label{eq:Amp}
\end{align}

From Eq. (\ref{eq:App}) and Eq. (\ref{eq:Amm}), we can obtain the Berry curvature as
\begin{align}
	\Omega_{z}^{+ +}
	=
	\left( \bm{\nabla} \times \bm{A}^{+ +} \right)_{z}
	=
	- \frac{r}{2} (\partial_{y} \varphi) \partial_{x} \left( \frac{\cos \theta}{\xi} \right)
	+
	\frac{r}{2} (\partial_{x} \varphi) \partial_{y} \left( \frac{\cos \theta}{\xi} \right),
	\label{eq:Omega_pp_z_1}
\end{align}
and using the relations 
\begin{align}
	&\frac{\partial}{\partial k_{x}}
	=
	\sin \theta \cos \varphi \frac{\partial}{\partial k}
	+
	\frac{\cos \theta \cos \varphi}{k} \frac{\partial}{\partial \theta}
	-
	\frac{\sin \varphi}{k \sin \theta} \frac{\partial}{\partial \varphi},
	\\
	&\frac{\partial}{\partial k_{y}}
	=
	\sin \theta \sin \varphi \frac{\partial}{\partial k}
	+
	\frac{\cos \theta \sin \varphi}{k} \frac{\partial}{\partial \theta}
	+
	\frac{\cos \varphi}{k \sin \theta} \frac{\partial}{\partial \varphi},
	\\
	&\frac{\partial}{\partial k_{z}}
	=
	\cos \theta \frac{\partial}{\partial k} - \frac{\sin \theta}{k} \frac{\partial}{\partial \theta},
\end{align}
we obtain
\begin{align}
	\Omega^{+ +}_{z}
	=
	\frac{r}{\xi^{3}} \frac{k_{z}}{2 k^{3}}.
	\label{eq:Omega_pp_z}
\end{align}
Similar calculation can be done and we obtain the Berry curvature as
\begin{align}
	\bm{\Omega}^{\sigma \sigma}
	=
	\sigma \frac{r}{\xi^{3}} \frac{\bm{k}}{k^{3}}.
	\label{eq:Omega}
\end{align}
We can see that Eq. (\ref{eq:Omega}) is same to the well-known result in the Weyl system 
when considering the isotropic case $r = \xi = 1$.

Similarly, we can calculate the ``inter-band Berry curvature'' as
\begin{align}
	\Omega^{+ -}_{x}
	=&
	\frac{1}{2 i}
	\left[ A^{+ -}_{y} A^{- +}_{z} - A^{+ -}_{z} A^{- +}_{y} \right]
	=
	\mathrm{Im} \left[ A^{+ -}_{y} A^{- +}_{z} \right]
	\notag \\
	=&
	\frac{r}{\xi^{3}} \frac{k_{x}}{4 k^{3}},
	\label{eq:Omega_pm_x}
	\\
	\Omega^{- +}_{x}
	=&
	- \frac{r}{\xi^{3}} \frac{k_{x}}{4 k^{3}},
	\label{eq:Omega_mp_x}
	\\
	\Omega^{+ -}_{y}
	=&
	\frac{r}{\xi^{3}} \frac{k_{y}}{4 k^{3}},
	\label{eq:Omega_pm_y}
	\\
	\Omega^{- +}_{y}
	=&
	-\frac{r}{\xi^{3}} \frac{k_{y}}{4 k^{3}}.
	\label{eq:Omega_mp_y}
\end{align}
Using the relation
\begin{align}
	\bm{m}_{n\bm{k}}
	=
	e
	\sum_{m}
	(\varepsilon_{m\bm{k}} - \varepsilon_{n\bm{k}}) \bm{\Omega}^{nm},
\end{align}
we obtain the orbital magnetic moment by 
\begin{align}
	&m^{x}_{\sigma}
	=
	- \sigma e \varepsilon_{\sigma}
	\frac{r}{\xi^{3}} \frac{k_{x}}{2 k^{3}},
	\label{eq:mx}
	\\
	&m^{y}_{\sigma}
	=
	- \sigma e \varepsilon_{\sigma}
	\frac{r}{\xi^{3}} \frac{k_{y}}{2 k^{3}}.
	\label{eq:my}
\end{align}

The velocities (derivatives of the energy dispersion) become
\begin{align}
	\partial_{x} \varepsilon_{\pm}
	=
	\pm \frac{v_{\mathrm{F}} k_{x}}{\xi k},
	\quad
	\partial_{y} \varepsilon_{\pm}
	=
	\pm \frac{v_{\mathrm{F}} k_{y}}{\xi k},
	\quad
	\partial_{z} \varepsilon_{\pm}
	=
	\pm \frac{v_{\mathrm{F}} r^{2} k_{x}}{\xi k},
\end{align}
then we obtain the NCTE Hall conductivity as
\begin{align}
	\sigma^{\mathrm{BC}}_{z}
	=&
	e^{2} \tau
	\sum_{\bm{k},\sigma} 
	(\varepsilon_{\sigma} - \mu)
	\left( - \frac{\partial f}{\partial \varepsilon} \right)_{\varepsilon_{\sigma}}
	\left[ 
		(\partial_{z} \varepsilon_{\sigma}) \Omega^{\sigma \sigma}_{z}
		-
		\frac{1}{2}
		\left\{ 
			(\partial_{x} \varepsilon_{\sigma}) \Omega^{\sigma \sigma}_{x}
			+
			(\partial_{y} \varepsilon_{\sigma}) \Omega^{\sigma \sigma}_{y}
		\right\}
	\right]
	\notag \\
	=&
	e^{2} \tau
	\sum_{\bm{k},\sigma}
	(\varepsilon_{\sigma} - \mu)
	\left( - \frac{\partial f}{\partial \varepsilon} \right)_{\varepsilon_{\sigma}}
	\frac{r v_{\mathrm{F}}}{2 \xi^{4} k^{4}}
	\left[ 
		r^{2} k_{z}^{2}
		-
		\frac{k_{x}^{2} + k_{y}^{2}}{2}
	\right]
	\notag \\
	=&
	\frac{e^{2} \tau}{(2 \pi)^{3}}
	\int_{0}^{\infty} \mathrm{d} k \int_{0}^{\pi} \mathrm{d} \theta \int_{0}^{2\pi} \mathrm{d} \varphi
	\sin \theta
	\sum_{\sigma}
	(\varepsilon_{\sigma} - \mu)
	\left( - \frac{\partial f}{\partial \varepsilon} \right)_{\varepsilon_{\sigma}}
	\frac{r v_{\mathrm{F}}}{2 \xi^{4}}
	\left[ 	
		r^{2} \cos^{2} \theta - \frac{1 - \cos^{2} \theta}{2}
	\right]
	\notag \\
	=&
	\frac{e^{2} \tau}{4 \pi^{2}}
	\int_{0}^{\pi} \mathrm{d} \theta
	\sin \theta
	\frac{r v_{\mathrm{F}}}{2 \xi^{4}}
	\left[ 	
		r^{2} \cos^{2} \theta - \frac{1 - \cos^{2} \theta}{2}
	\right]
	\int_{0}^{\infty} \mathrm{d} k 
	\sum_{\sigma}
	(\varepsilon_{\sigma} - \mu)
	\left( - \frac{\partial f}{\partial \varepsilon} \right)_{\varepsilon_{\sigma}}
	\notag \\
	=&
	\frac{e^{2} \tau}{4 \pi^{2}}
	\int_{0}^{\pi} \mathrm{d} \theta
	\sin \theta
	\frac{r v_{\mathrm{F}}}{2 \xi^{4}}
	\left[ 	
		r^{2} \cos^{2} \theta - \frac{1 - \cos^{2} \theta}{2}
	\right]
	\frac{1}{v_{\mathrm{F}} \xi} 
	\int_{-\infty}^{\infty} \mathrm{d} k 
	\left( v_{\mathrm{F}} \xi k - \mu \right)
	\left( - \frac{\partial f(v_{\mathrm{F}} \xi k)}{\partial k} \right)
	\notag \\
	=& 0,
	\label{eq:sigma_BC}
	\\
	\sigma^{\mathrm{OM}}_{z}
	=&
	- e \tau
	\sum_{\bm{k},\sigma}
	\partial_{\varepsilon} 
	\left\{ (\varepsilon - \mu) \left( - \frac{\partial f}{\partial \varepsilon} \right) \right\}_{\varepsilon_{\sigma}}
	\left[ 
		(\partial_{x} \varepsilon_{\sigma}) m^{x}_{\sigma}
		+
		(\partial_{y} \varepsilon_{\sigma}) m^{y}_{\sigma}
	\right]
	\notag \\
	=&
	e^{2} \tau
	\sum_{\bm{k},\sigma}
	\partial_{\varepsilon} 
	\left\{ (\varepsilon - \mu) \left( - \frac{\partial f}{\partial \varepsilon} \right) \right\}_{\varepsilon_{\sigma}}
	\varepsilon_{\sigma}
	\frac{r v_{\mathrm{F}}}{2 \xi^{4} k^{4}}
    \frac{k_{x}^{2} + k_{y}^{2}}{2}
	\notag \\
	=&
	\frac{e^{2} \tau}{4 \pi}
	\int_{0}^{\infty} \mathrm{d} k 
	\int_{0}^{\pi} \mathrm{d} \theta
	\sin \theta
	\sum_{\sigma}
	\partial_{\varepsilon}
	\left\{ 
		(\varepsilon - \mu) \left( - \frac{\partial f}{\partial \varepsilon} \right)
	\right\}_{\varepsilon_{\sigma}}
	\varepsilon_{\sigma}
	\frac{r v_{\mathrm{F}}}{3 \xi^{4}}
	\frac{1 - \cos^{2} \theta}{2}
	\notag \\
	=&
	\frac{e^{2} \tau}{4 \pi}
	\int_{0}^{\pi} \mathrm{d} \theta
	\sin \theta
	\frac{r v_{\mathrm{F}}}{3 \xi^{4}}
	\frac{1 - \cos^{2} \theta}{2}
	\int_{- \infty}^{\infty} \mathrm{d} k 
	\partial_{\varepsilon}
	\left\{ 
		(\varepsilon - \mu) \left( - \frac{\partial f}{\partial \varepsilon} \right)
	\right\}_{\varepsilon=v_{\mathrm{F}} \xi k}
	v_{\mathrm{F}} \xi k
	\notag \\
	=&
	\frac{e^{2} \tau}{4 \pi}
	\int_{0}^{\pi} \mathrm{d} \theta
	\sin \theta
	\frac{r v_{\mathrm{F}}}{3 \xi^{4}}
	\frac{1 - \cos^{2} \theta}{2}
	\frac{1}{v_{\mathrm{F}} \xi}
	\int_{- \infty}^{\infty} \mathrm{d} k 
	\partial_{k}
	\left\{ 
		(v_{\mathrm{F}} \xi k - \mu) \left( - \frac{\partial f(v_{\mathrm{F}} \xi k)}{\partial k} \right)
	\right\}
	k
	\notag \\
	=&
	-
	\frac{e^{2} \tau}{4 \pi}
	\int_{0}^{\pi} \mathrm{d} \theta
	\sin \theta
	\frac{r v_{\mathrm{F}}}{3 \xi^{4}}
	\frac{1 - \cos^{2} \theta}{2}
	\frac{1}{v_{\mathrm{F}} \xi}
	\int_{- \infty}^{\infty} \mathrm{d} k 
	(v_{\mathrm{F}} \xi k - \mu) \left( - \frac{\partial f(v_{\mathrm{F}} \xi k)}{\partial k} \right)
	\notag \\
	=&
	0.
	\label{eq:sigma_OM}
\end{align}

%%%%%%%%%% -------------------------------------------------- %%%%%%%%%%

\twocolumngrid

%\bibliography{nonlinear}
%\clearpage

\bibliography{nonlinear}
\clearpage
\end{document}